\documentclass[superscriptaddress, preprint, preprintnumbers]{revtex4-2}
\usepackage{graphicx}

\usepackage{amssymb}
\usepackage{amsmath}
\usepackage{amsthm}
\usepackage{amsfonts}
\usepackage{xfrac}
\usepackage{wasysym}
\usepackage{textcomp}
\usepackage{bbm}

\usepackage{braket}
\usepackage{mhchem}
\usepackage{epstopdf} 

\usepackage{hyperref}
\usepackage{array}
\usepackage[utf8]{inputenc}
\usepackage[table]{xcolor}

\newcommand{\figSize}{0.8}

\begin{document}
\preprint{LA-UR-21-20802}

\author{Stefano Marin}
\email{stmarin@umich.edu}
\affiliation{Department of Nuclear Engineering and Radiological Sciences, University of Michigan, Ann Arbor, MI 48109, USA}

\author{M. Stephan Okar}
\affiliation{Department of Nuclear Engineering and Radiological Sciences, University of Michigan, Ann Arbor, MI 48109, USA}

\author{Eoin P. Sansevero}
\affiliation{Department of Nuclear Engineering and Radiological Sciences, University of Michigan, Ann Arbor, MI 48109, USA}

\author{Isabel E. Hernandez}
\affiliation{Department of Nuclear Engineering and Radiological Sciences, University of Michigan, Ann Arbor, MI 48109, USA}

\author{Catherine A. Ballard}
\affiliation{Department of Nuclear Engineering and Radiological Sciences, University of Michigan, Ann Arbor, MI 48109, USA}

\author{Ramona Vogt}
\affiliation{Nuclear and Chemical Sciences Division, Lawrence Livermore National Laboratory, Livermore, CA 94550, USA}
\affiliation{Department of Physics and Astronomy Department, University of California, Davis, CA 95616, USA}

\author{J\o rgen Randrup}
\affiliation{Nuclear Science Division, Lawrence Berkeley National Laboratory, Berkeley, CA 94720, USA}

\author{Patrick Talou}
\affiliation{Computational Physics Division, Los Alamos National Laboratory, Los Alamos, NM 87545, USA}

\author{Amy E. Lovell}
\affiliation{Theoretical Physics Division, Los Alamos National Laboratory, Los Alamos, NM 87545, USA}

\author{Ionel Stetcu}
\affiliation{Theoretical Physics Division, Los Alamos National Laboratory, Los Alamos, NM 87545, USA}

\author{Olivier Serot}
\affiliation{CEA, DES, IRESNE, DER, SPRC, Physics Studies Laboratory, Cadarache, F-13108 Saint-Paul-lès-Durance, France}

\author{Olivier Litaize}
\affiliation{CEA, DES, IRESNE, DER, SPRC, Physics Studies Laboratory, Cadarache, F-13108 Saint-Paul-lès-Durance, France}

\author{Abdelhazize Chebboubi}
\affiliation{CEA, DES, IRESNE, DER, SPRC, Physics Studies Laboratory, Cadarache, F-13108 Saint-Paul-lès-Durance, France}

\author{Shaun D. Clarke}
\affiliation{Department of Nuclear Engineering and Radiological Sciences, University of Michigan, Ann Arbor, MI 48109, USA}

\author{Vladimir A. Protopopescu}
\affiliation{Oak Ridge National Laboratory, Oak Ridge, TN 37830, USA}

\author{Sara A. Pozzi}
\affiliation{Department of Nuclear Engineering and Radiological Sciences, University of Michigan, Ann Arbor, MI 48109, USA}
\affiliation{Department of Physics, University of Michigan, Ann Arbor, MI 48109, USA}

\title{Structure in the Event-by-Event Energy-Dependent Neutron-Gamma Multiplicity Correlations in \ce{^{252}Cf}(sf)}
\date{\today}

\begin{abstract}
The emission of neutrons and gamma rays by fission fragments reveal important information about the properties of fragments immediately following scission. The initial fragment properties, correlations between fragments, and emission competition give rise to correlations in neutron-gamma emission. Neutron-gamma correlations are important in nonproliferation applications because the characterization of fissionable samples relies on the identification of signatures in the measured radiation. Furthermore, recent theoretical and experimental advances have proposed to explain the mechanism of angular momentum generation in fission. In this paper, we present a novel analysis method of neutrons and gamma rays emitted by fission fragments that allows us to discern structure in the observed correlations. We have analyzed data collected on \ce{^{252}Cf}(sf) at the Chi-Nu array at the Los Alamos Neutron Science Center. Through our analysis of the energy-differential neutron-gamma multiplicity covariance, we have observed enhanced neutron-gamma correlations, corresponding to rotational band gamma-ray transitions, at gamma-ray energies of $0.7$ and $1.2$ MeV. To shed light on the origin of this structure, we compare the experimental data with the predictions of three model calculations. The origin of the observed correlation structure is understood in terms of a positive spin-energy correlation in the generation of angular momentum in fission. 

\end{abstract}


\maketitle

\section{Introduction}

Heavy atomic nuclei can decay by fission, \textit{i.e.},  separating into two or more large nuclear fragments. The complexities of the strong nuclear forces and the electromagnetic interactions between hundreds of nucleons lead to a variety of exit channels for the fission process, resulting in stochastic distributions of both masses and kinetic energies of the produced fission fragment~\cite{Brosa1990}. Large correlated variations exist also in the excitation energies and angular momenta of the fragments~\cite{Wilhelmy1972, Nifenecker1972, Schmidt2011}, as well as in the emission of neutrons and gamma rays~\cite{Talou2018}. The initial conditions of fission fragments cannot be directly probed due in part to the short time scales of fragment de-excitation. 
To study the condition of the fragments immediately after scission, we study the emission of neutrons and gamma rays, since their emission is highly correlated with the initial fragment conditions~\cite{Wilhelmy1972, Hoffman1964}. Particularly important open questions in fission modeling are the generation of angular momentum at scission~\cite{SchmidFabian1988, Nifenecker1972, Brosa1985} and the sharing of excitation energy between fragments~\cite{Schmidt2011, Signarbieux1972}. 

In a previous publication \cite{Marin2020}, we surveyed the results of several earlier publications on event-by-event neutron-gamma multiplicity correlations in \ce{^{252}Cf}(sf). 
After developing and applying analytic unfolding techniques to the experimental data, we found these data to be in qualitative agreement with one another and with the predictions of model calculations. 
Specifically, the results indicate the existence of significant negative event-by-event correlations between the neutron-gamma multiplicities. However, these experimental results were affected by energy-dependent systematic biases. 

Understanding neutron-gamma correlations is of great importance in nonproliferation applications. The negative neutron-gamma covariance represents a signature of fissioning special nuclear material. In this work, we find that the energy-dependent covariance contains a more detailed signature of the fission reaction, which changes drastically in magnitude and sign across the examined energies. Furthermore, the characterization of the energy-dependence of neutron-gamma correlations is an important component in addressing systematic biases related to other observables. 

The goal of this paper is to further develop our previous analysis of neutron-gamma correlations by studying energy-dependent correlations in the spontaneous fission of \ce{^{252} Cf}. We find that the structure of the correlations observed in the experiment can be explained by known and predicted physical mechanisms and indicates the existence of an energy-dependence in the generation of angular momentum in fission. 

Section~\ref{sec:the} presents a brief overview of the physical mechanisms that give rise to correlations in fission.
Section~\ref{sec:ana} presents the mathematical formalism required to extract and quantify the predicted correlations. 
Section~\ref{sec:exp} reports the experimental setup and the techniques used in experiment. In Section~\ref{sec:dis}, we describe the most prominent features of the neutron-gamma correlations and compare them with model calculations. Lastly, in Section~\ref{sec:conc}, we draw our conclusions on the origin of the observed correlations. 

\section{Sources of Correlations}
\label{sec:the}

\subsubsection*{Neutron and Gamma-Ray Emission}

\begin{figure}[!htb]
\centering
\includegraphics[width=\figSize \linewidth]{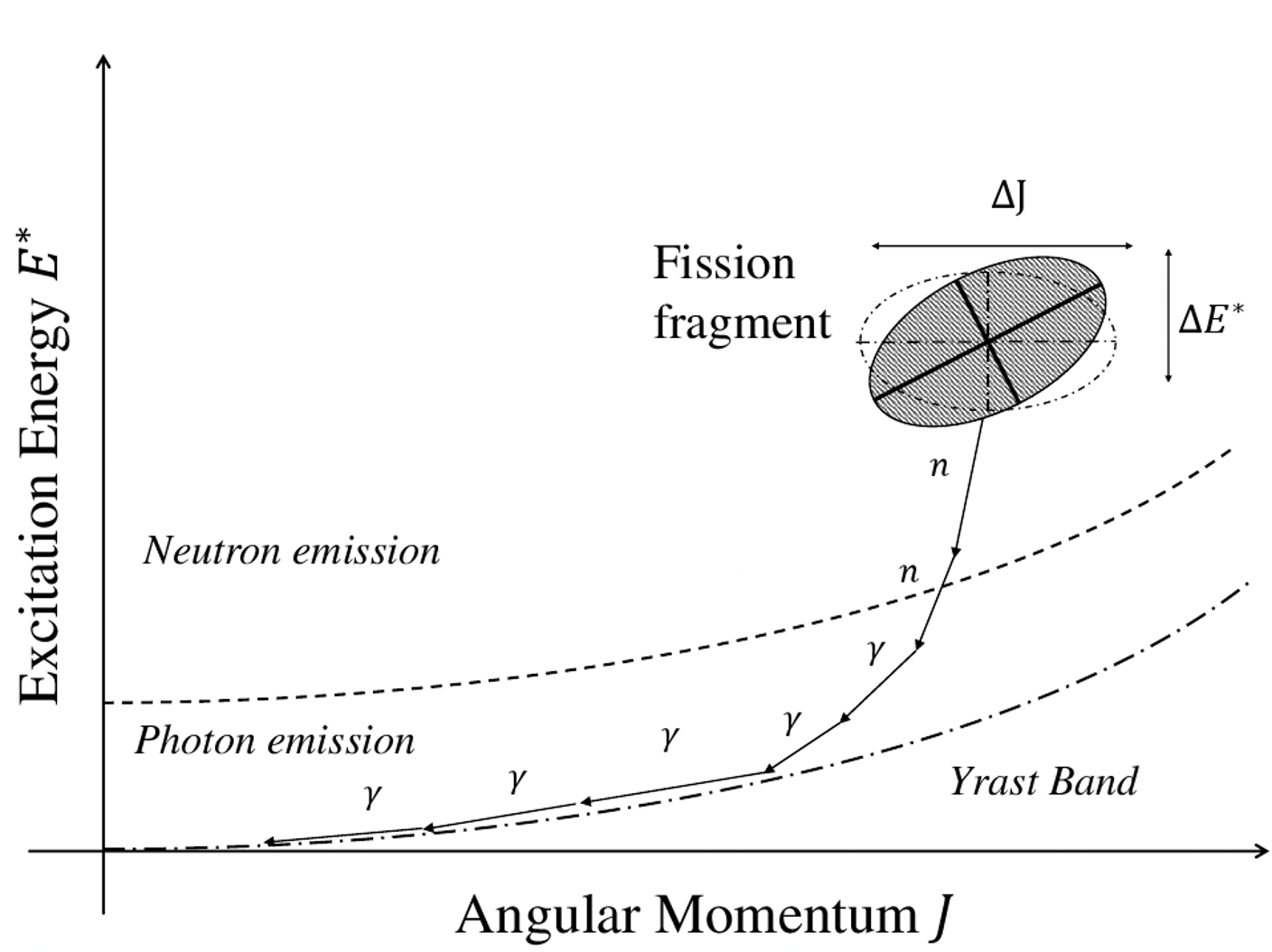}
\caption{Schematic representation of neutron and gamma-ray emission from fission fragments. The neutron separation energy is indicated by a dashed line, and the Yrast rotational band is indicated as a dot-dashed line~\cite{Thomas1967}.} 
\label{fig:deExc}
\end{figure}

Following scission, fission fragments promptly decay by neutron emission, if energetically possible, followed by gamma-ray emission. Two modes of gamma-ray emission are usually distinguished \cite{Serot2010, Talou2018}: statistical gamma rays in the continuum at higher intrinsic excitation energies (\textsc{stat}), and band-transition gamma rays in the discrete-level region (\textsc{disc}). A pictorial diagram, inspired by Ref.~\cite{Serot2010}, is shown in Fig.~\ref{fig:deExc}. We have explicitly drawn a tilted initial distribution of the initial fragment conditions to highlight suspected spin-energy correlations in the formation of fission fragments~\cite{Nifenecker1972, SchmidFabian1988}. 

It is worthwhile noting the different roles played by neutron and gamma-ray emission in fragment de-excitation. At scission, the total available excitation energy is partitioned between the two fragments in the form of intrinsic energy, shape deformation, and collective motion~\cite{Schmidt2011}. Neutron and gamma-ray emission takes place from fully accelerated fragments, giving enough time for the nuclear deformation to relax into excitation energy. Neutrons, due to their binding energies, dissipate the largest portion of the fragment intrinsic energy, \textit{i.e.}, the energy not stored in collective rotational motion of nucleons. \textsc{Stat} gamma rays dissipate the remaining intrinsic excitation energy until a rotational band is reached. \textsc{disc} gamma rays dissipate the remaining rotational energy and angular momentum. The emission of \textsc{disc} gamma rays is regulated most strongly by the feeding of rotational band levels, which in turn depends on the initial angular momenta of the fragment. 

In the remainder of this section we will investigate the sources of possible neutron-gamma correlations. Positive neutron-gamma multiplicity correlations, in the present context, indicate that the emission of each neutron makes it more likely to emit a gamma ray. Negative correlations represent the opposite.

\subsubsection*{Decay Competition}

The most direct source of correlations is the competition between neutron and gamma-ray emission. In theory, this competition could result in very strong negative correlations between the neutron and gamma-ray multiplicity. In practice, the two decay modes can be treated as independent apart from a narrow intrinsic energy range near the neutron separation energy~\cite{Thomas1967, Spyrou2016}. The gamma rays competing with the neutrons are \textsc{stat}. Furthermore, because the fragments are less excited close to the neutron separation energy~\cite{Blatt1952}, we expect these negative correlations to appear at low  neutron energies. 

\subsubsection*{Resource Competition}

Another significant source of neutron-gamma competition is that introduced by the finite excitation energy and angular momentum of the fragments. In fact, given the same excitation energy $E^*$, a fragment that dissipates more of its intrinsic energy through neutron evaporation will have less energy to dissipate through \textsc{stat} gamma-ray emission. The competition over the intrinsic energy of the fragments would result in overall negative correlations between neutrons and \textsc{stat} gamma rays.

Competition related to the fragment angular momentum $J$ are more complicated, given the vector nature of the angular momentum. If neutron emission is found to remove significant quantities of angular momentum, we expect negative correlations between neutrons and \textsc{disc} gamma rays, since these are the gamma rays mostly affected by the angular momenta of the fragments. 

\subsubsection*{Excitation Energy and Angular Momentum}

Correlations between neutrons and gamma rays are expected if their emission depends on the same fragment properties. For example, we know that more energetic fission fragments will emit more energetic neutrons and \textsc{stat} gamma rays. Therefore we expect the spectra of neutrons and gamma rays to receive contributions from their shared dependence on the fragment temperature \cite{Blatt1952}. 

The interesting and still unclear mechanism for angular momentum generation~\cite{Wilhelmy1972, Bonneau2007, Rakapoulos2018, Wilson2021, Randrup2021} can lead to correlated behavior in neutron-gamma emission. In fact, the intrinsic excitation energy is strongly linked to the neutron multiplicity~\cite{Gook2014}, while the emission of \textsc{disc} gamma rays is regulated by the angular momentum~\cite{Nifenecker1972, Vogt2021}. Thus, if $E^*$ and $J$ are positively correlated, \textit{i.e.}, positive spin-energy correlations, as phenomenological and theoretical models predict~\cite{Bonneau2007, Bertsch2015, Vogt2014b}, we also expect positive correlations between neutrons and \textsc{disc} gamma rays.

These correlations will be especially important in this work, since \textsc{disc} gamma rays have similar energies for many different probable fragments in fission, and can thus give rise to correlated structure. Experimentally, spin-energy correlations have not been conclusively proven. In fact their existence has been disputed by some~\cite{Wilhelmy1972, SchmidFabian1988}, and supported by others~\cite{Nifenecker1972, Chebboubi2017}. See also Refs.~\cite{Frehaut1983, Wang2016, Thulliez2019}. Clearly, more experimental evidence regarding possible spin-energy correlations is needed.

We note that gamma-ray emission is correlated with the properties of the post-evaporation fission fragments. Therefore, the energy-spin correlations can be counterbalanced by the dissipation of the fragment energy and angular momentum during neutron emission. 

\subsubsection*{Inter-Fragment Correlations}

While all of the sources of correlations discussed above are related to neutrons and gamma rays from the same fragment, it is also possible to have correlations between emissions from opposite fragments. The strongest source of inter-fragment correlations are those related to the initial excitation energy and angular momentum of the fragments. If the fragments are assumed to have strongly correlated initial conditions, we expect the correlations between neutrons and gamma rays originating from opposite fragments (inter-fragment) to be of the same order of magnitude as those between particles originating in the same fragment (intra-fragment). 

Because inter-fragment correlations do not suffer from internal decay competition or resource competition, we expect inter-fragment correlations to lack negative sources. Stated in an equivalent way, we expect inter-fragment correlations to be positive overall. 

A noteworthy special case is mentioned here. If the initial fragment conditions are strongly correlated to one another but one of the  fragments emits significantly fewer neutrons than the other, we expect inter-fragment correlations to be large. In fact, the angular momentum of the fragment emitting fewer neutrons would be less changed by neutron emission.

\subsubsection*{Summary}

We summarize the discussion of the possible origins of neutron-gamma correlations in Table~\ref{tab:sumCorr}. For each source discussed, we specify whether we expect it to be introducing positive or negative correlations. We also distinguish, in the $\gamma$-type column, whether the correlations arise between neutrons and \textsc{stat} gamma rays or \textsc{disc} gamma rays. Lastly, we indicate whether these correlations can also arise from inter-fragment correlations. 

\begin{table}
\caption{\label{tab:sumCorr} Summary of the expected sources of neutron-gamma correlations}
\begin{tabular}{|l||l|l|l|}
\hline
Source & Sign of Correlation & $\gamma $ - type & Inter-Fragment \\
\hline
\hline
Decay competition & Negative & \textsc{stat} & No\\
\hline
Energy competition & Negative & \textsc{stat} & No\\
\hline
Angular momentum competition & Negative & \textsc{disc} & No\\
\hline
Fragment temperature & Overall neutral & \textsc{stat} & Yes\\
\hline
Spin-Energy correlation & Positive & \textsc{disc} & Yes \\ 
\hline
\end{tabular}
\end{table}

\section{Analysis}
\label{sec:ana}

Having discussed why and where we expect to see correlations, we need to quantify their magnitude. In this section we discuss a new type of statistical analysis of radiation data, the normalized differential multiplicity covariance. 

\subsubsection*{Covariance}

Let us define the event-by-event neutron and gamma multiplicities, $N_n$ and $N_\gamma$. We investigate the correlations in the neutron-gamma emission by analyzing the multiplicity covariance $\text{cov}(N_n, N_\gamma)$. This quantity describes the linear correlations in the event-by-event fluctuations of the neutron and gamma multiplicities with respect to the mean values. We have already shown in our previous study~\cite{Marin2020} that the neutron-. \textsc{disc} multiplicity covariance is negative. Decay competition has a a negligible effect on the correlations integrated over all energies. Using Table~\ref{tab:sumCorr}, we conclude that energy and angular momentum competition dominate over the other sources of correlations. To investigate the other sources, we analyze the dependence of the neutron-gamma covariance on the energies of the outgoing particles. 

\subsubsection*{Differentials}

The energy-dependent covariance is introduced as the covariance between the neutron and gamma-ray event-by-event spectra, 
\begin{equation}
    \text{cov} \left( \frac{d N_n}{d E_n}, \frac{d N_\gamma}{d E_\gamma} \right) = \frac{\partial^2 \text{cov} \left (N_n, N_\gamma \right)}{\partial E_n \partial E_\gamma}, 
\label{eq:covDiff}
\end{equation}
where $E_n$ and $E_\gamma$ are the neutron and gamma-ray energies. The arguments of the covariance in the left-hand side of  Eq~\eqref{eq:covDiff} are the event-by-event multiplicities differentiated by spectra, \textit{i.e.}, the number of particles emitted with a specified energy, listed for each fission event. The right-hand side of Eq.~\eqref{eq:covDiff} then follows from the bilinearity of the covariance operator. The energy-differentiated covariance is then interpreted as the correlations between neutron and gamma rays multiplicities, if only neutrons and gamma rays of energies $E_n$ and $E_\gamma$ are counted. The differentiated covariance allows us to investigate how the emission of neutrons of a specified energy affects the emission of gamma rays with another specified energy. For sufficiently narrow energy bins, the event-by-event multiplicity in each bin also tends to zero. For sufficiently small multiplicities, the correlations approach linearity and are described appropriately by the covariance.

\subsubsection*{Covariance Scaling}

While the differential covariance describes the correlations between neutrons and gamma rays, it also scales with the multiplicity in each energy bin. To compare the strength of the correlations across energies, we normalize the covariance by the independent spectra, \textit{i.e.}, 
\begin{equation}
    C_{E_n, E_\gamma } = \frac{\partial^2 \text{cov} \left (N_n, N_\gamma \right)}{\partial E_n \partial E_\gamma} \left( \frac{d \langle N_n \rangle }{d E_n} \frac{d \langle N_\gamma \rangle }{d E_\gamma} \right)^{-1}. 
\end{equation}
With this choice of normalization, $C_{E_n, E_\gamma }$ satisfies
\begin{equation}
    C_{E_n, E_\gamma } \geq -1,
\end{equation}
where the equality holds only in the case where the emission of a neutron of energy $E_n$ completely precludes the emission of a gamma ray of energy $E_\gamma$, and vice versa.

\section{Experiment}
\label{sec:exp}

\subsubsection*{Experiment}

We analyze the data of Marcath \textit{et al.} ~\cite{Marcath2018} collected using the Chi-Nu array at the Los Alamos neutron science center (LANSCE). We have analyzed $1.7\times10^{10}$ \ce{^{252}Cf} spontaneous fission events, each tagged by an ionization chamber fabricated at Oak Ridge National Laboratory~\cite{Kirsch2017}. Due to hardware limitations, only $42$ of the $54$ EJ-309 organic scintillation detectors were active during the measurement. Each detector was placed at a distance of approximately $1$ m from the fission source. Coincidences between the EJ-309 detectors and the fission chamber are employed to measure the neutron and gamma-ray multiplicities, $D_n$ and $D_\gamma$ respectively, detected in each event. For more information on this experiment, see Refs~\cite{Marcath2018, Schuster2019, Marin2020}. 

\subsubsection*{Simulation and Response}

\begin{figure}[!htb]
\centering
\includegraphics[width=\figSize \linewidth]{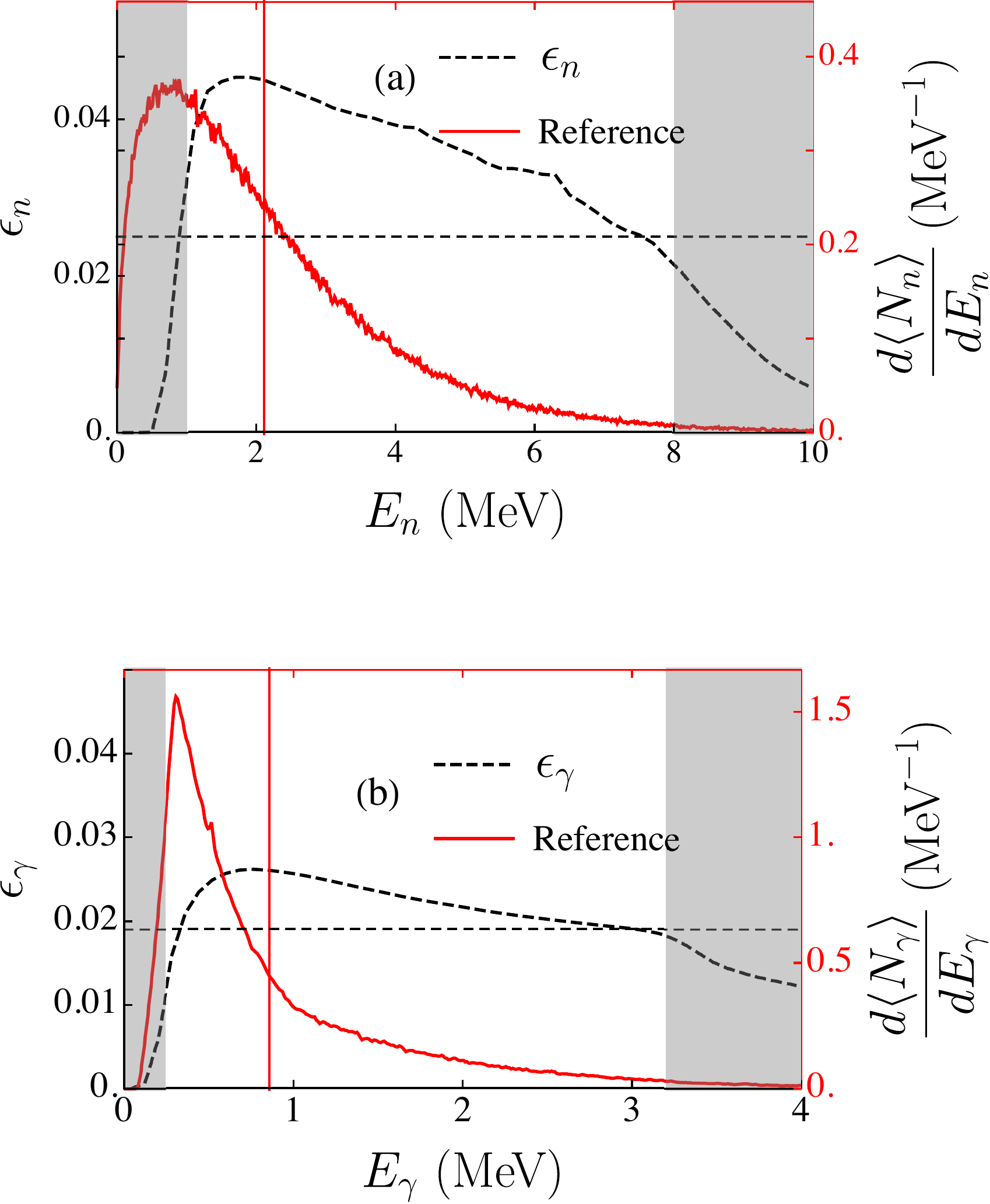}
\caption{Simulated energy response of the Chi-Nu array to (a) neutrons and (b) gamma rays. The solid red line, referencing the right-hand axis, shows the evaluated spectrum included in the standard MCNPX-PoliMi californium source \cite{Pozzi2003, Pozzi2012a}. The vertical red and horizontal dashed black lines represent respectively the mean emission energy and mean system efficiency. The grey bands indicate the regions outside the energy acceptance of the system.} 
\label{fig:energyResp}
\end{figure}

We performed a high-fidelity simulation of the system response using MCNPX-PoliMi~\cite{Pozzi2003, Pozzi2012a}, a Monte Carlo-based radiation transport code. In the simulation, the response of each detector is individually modeled, taking into account any differences in distance to source, light output, and calibration. We introduce a small perturbation in the light outputs and the array geometry in the simulations to better reproduce the reference evaluated neutron spectrum after radiation transport. The parameters of the light output response for each detector, modeled analytically as exponential functions \cite{Norsworthy2017}, are slightly perturbed from the literature values to better fit the reference spectrum. Clearly, the unfolding procedure is systematically biased to reproduce the known \ce{^{252}Cf}(sf) neutron spectrum. No corrections are applied to reproduce the gamma-ray spectrum. The analysis of the simulated system response is presented in Fig.~\ref{fig:energyResp}. The absolute detection efficiencies, $\epsilon_n$ and $\epsilon_\gamma$, are not uniform across the emitted particle spectra. 

% \begin{figure}[h]
% \centering
% \includegraphics[width=0.45\linewidth]{figAngEffBin.eps}
% \caption{Angular response of the ChiNu array to neutron-photon pairs. In the upper quadrants, each pair of detectors is plotted at angle that they make with one another, and at a distance from the center proportional to the uncorrelated detection rate from the pair. In the lower half, the pairs are binned in $0.1$ rad bins, with the size of each marker proportional to the bin population, and the lines representing the standard deviations of the bin samples.}
% \label{fig:angleResp}
% \end{figure}

\subsubsection*{Unfolding Matrices}

In experiment, we do not have access to the emitted multiplicities, $N_n$ and $N_\gamma$, or the emitted neutron and gamma-ray energies $E_n$ and $E_\gamma$. Rather, we have access to detected multiplicities $D_n$ and $D_\gamma$, differentiated with respect to time of flight energy $T_n$ for neutrons, and energy deposited by gamma rays $F_\gamma$. Due to the large distance between detectors and source, time of flight leads to accurate assignment of the energy up to $E_n \approx 7$ MeV. However, because gamma rays interact through Compton scattering with the liquid organic scintillator, only a fraction of the energy is deposited in each gamma-ray interaction. 

In our simulations, we calculate the matrix $dE_n/dT_n$ describing, for each neutron emitted with energy $E_n$, the probability of assigning it an energy of $T_n$. Similarly, we calculate $dE_\gamma/dF_\gamma$ for the probability of a gamma ray of energy $E_\gamma$ to deposit energy $F_\gamma$. Energy and time resolution, determined from the experimental data, have been applied in simulation to smooth out all calculated results. 

The unfolding of the measured neutron-gamma correlations and spectra is performed using the response matrices and the measured neutron-gamma correlations, 
\begin{equation}
    \frac{\partial^2 \text{cov}(N_n, N_\gamma)}{\partial E_n \partial E_\gamma} = \frac{ \partial^2 \text{cov}(D_n, D_\gamma) }{\partial T_n \partial F_\gamma} \left( \epsilon_n \frac{dE_n}{dT_n} \right)^{-1} \left( \epsilon_\gamma \frac{dE_\gamma}{dF_\gamma} \right)^{-1}\ , 
    \label{eq:unfEM}
\end{equation}
\begin{subequations}
  \begin{equation}
    \frac{d \langle N_n \rangle }{d E_n} =  \frac{d \langle D_n \rangle }{d T_n} \left(\epsilon_n \frac{dE_n}{dT_n} \right)^{-1}
    \end{equation}
    \begin{equation}
    \frac{d \langle N_\gamma \rangle }{d E_\gamma} =  \frac{d \langle D_\gamma \rangle }{d F_\gamma} \left(\epsilon_\gamma \frac{dE_\gamma}{dF_\gamma} \right)^{-1} \ .  
  \end{equation}
\label{eq:unfSP}
\end{subequations}

\subsubsection*{Regularization}

\begin{figure}[!htb]
\centering
\includegraphics[width=\figSize \linewidth]{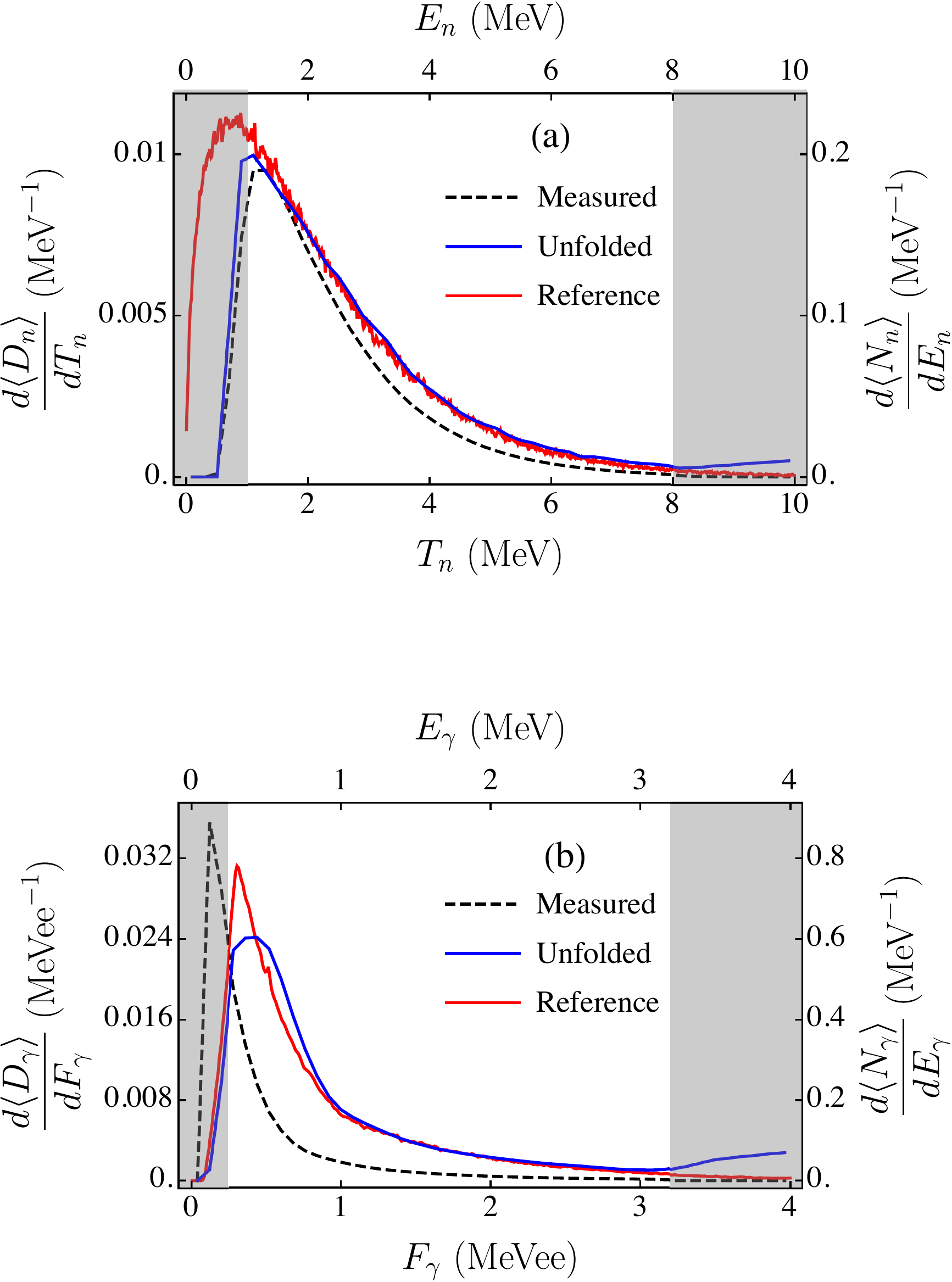}
\caption{Measured spectra and unfolded spectra after Tikhonov regularization. The same reference spectra given in Fig.~\ref{fig:energyResp} are also shown in red.} 
\label{fig:specUnf}
\end{figure}

To reduce the effects of noise in the unfolded distributions, we have applied Tikhonov regularization \cite{IllPosedMorozov}. In this unfolding procedure, regularization parameters $\alpha_n$ and $\alpha_\gamma$ are introduced. The unfolding procedure is optimized and validated on the measured spectra, where we have determined the values of and uncertainties on $\alpha_n$ and $\alpha_\gamma$. By comparison with the evaluated spectra, we have determined the effective acceptance regions in the unfolded distributions, $1.2 \leq E_n \leq 8 $ MeV and $0.22 \leq E_\gamma \leq 3.2 $ MeV. These acceptance regions, indicated by the areas between the shaded bands in Fig.~\ref{fig:energyResp}, cover $66\% $ and $78\%$ of the neutron and gamma-ray spectra, respectively. The results of the spectral unfolding procedure are shown in  Fig.~\ref{fig:specUnf}. 

\section{Results}
\label{sec:dis}

\begin{figure}[!htb]
\centering
\includegraphics[width=\linewidth]{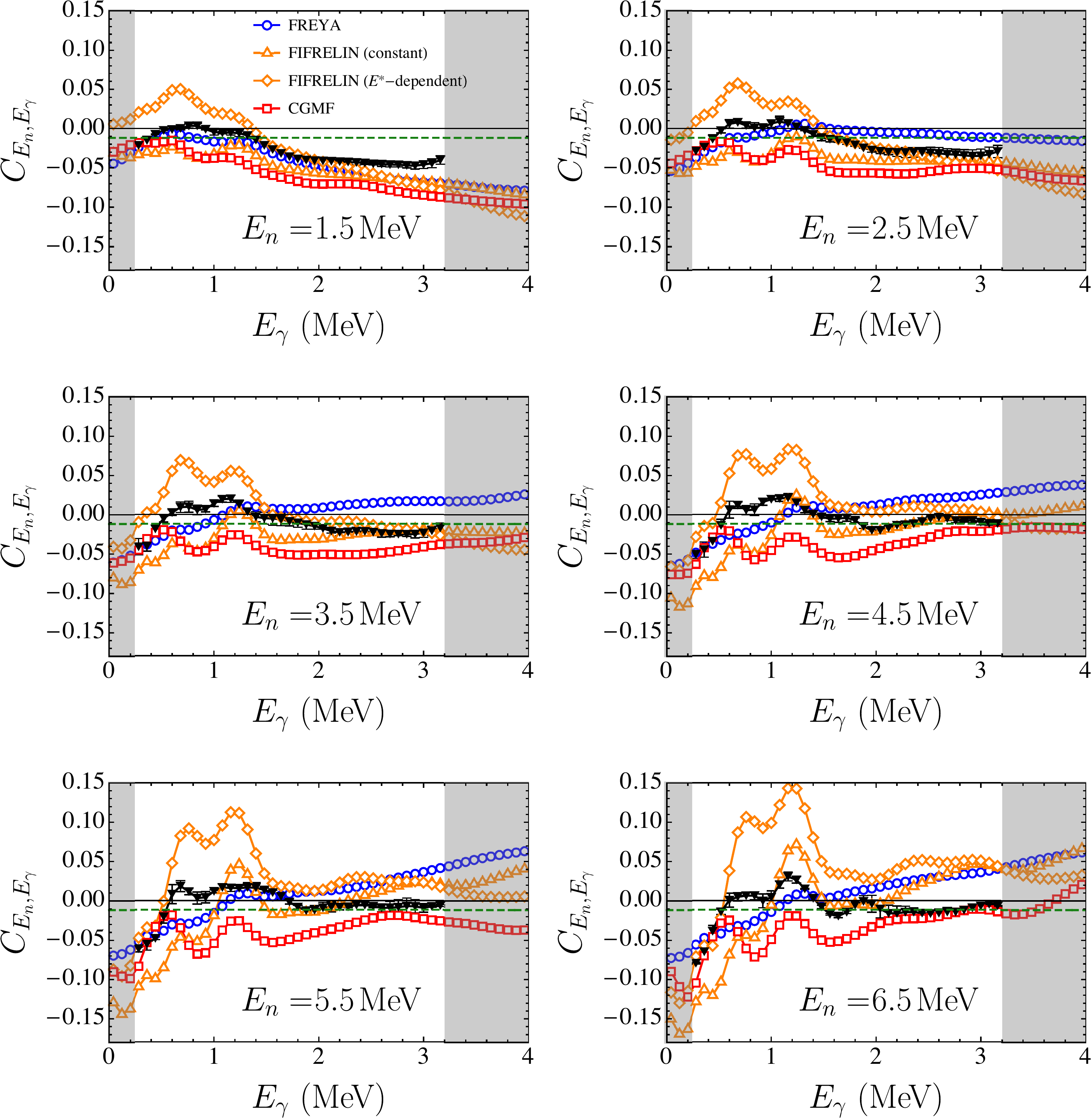}
\caption{Partial derivative analysis of the correlations. In each panel, the fixed neutron energy is indicated, and a green band represents the average of the experiment over all energies. The experimental data are shown in black, the error bars are given based on variations within neighboring $E_n$ slices. The horizontal green line represents the correlations averaged across all energies.}
\label{fig:partAn}
\end{figure}

The unfolding procedure introduced in the last section is applied to the experimental data. We have analyzed the data using energy bins of width $0.2$ MeV for $E_n$ and $0.08$ MeV for $E_\gamma$. We present our experimental determination of $C_{E_n, E_\gamma}$ in Fig.~\ref{fig:partAn}. We have included systematic errors associated with the unfolding by including experimental curves above and below the indicated fixed $E_n$. Systematic biases associated with background, particle misclassification, detector blinding, and inelastic gamma-ray production are removed according to the procedure outlined in Ref.~\cite{Marin2020}. To facilitate the visualization of the data, we have included a horizontal dashed line indicating the average of the correlations across all $E_n$ and $E_\gamma$.

We observe structure in the form of two enhancements at $E_\gamma \approx 0.7$ and $1.2$ MeV. A downslope develops at the lowest energies, extending down to the rejection region at $< 0.3$  MeV. Higher-energy correlation enhancements start to appear with increasingly higher gates on neutron outgoing energies, notably at $E_\gamma \approx 1.7$ MeV. 

The experimental results are compared to the \texttt{CGMF}~\cite{Talou2018, Talou2020}, \texttt{FIFRELIN}~\cite{Serot2010, Thulliez2019, Litaize2015}, and \texttt{FREYA}~\cite{Vogt2009, Verbeke2018} model calculations. The release versions of \texttt{CGMF} and \texttt{FREYA} are used here. \texttt{FIFRELIN} results are shown for an energy independent (constant) and an energy dependent ($ E^* $-dependent) model for the generation of angular momentum in the fragments~\cite{Thulliez2016}. A Gaussian smoothing informed by the energy resolution of the system is applied to all model calculations to be consistent with the data analysis. We note that the goal of the current analysis is not to compare any individual model calculations to the data, but rather to determine which features of the data differ from all the model calculations and can therefore offer insights into the fission mechanism. The feature of the correlations we focus on in this work are the correlations enhancements at $E_\gamma = 0.7$ and $1.2$ MeV. 

\subsubsection*{Enhanced Correlation Structure}

The two correlation enhancements at $0.7$ and $1.2$ MeV appears to arise from \textsc{disc} gamma rays positively correlating to the emission of neutrons. The gamma rays giving rise to these enhancements cannot be assigned to a single fragment level scheme, but are rather emergent behaviors from the combination of many transitions of similar energies. We note that much more work, both experimental and theoretical, is needed in order to quantify the spin-energy correlations. 

The enhancement at $1.2$ MeV corresponds to transitions from quasi-spherical nuclei near \ce{^{132} Sn}. 
The enhancement at $0.7$ MeV receives contributions from both light and heavy fragments, and it is due to electric quadrupole Yrast transitions with initial spins $4 \hbar$, $6 \hbar$, and $8 \hbar$. The predicted valley at $E_\gamma < 0.4 $ MeV is due to competition between neutron emission and \textsc{disc} gamma rays. 

% The presence of positive correlations at high $E_\gamma$ has several origins. We note here that we are discussing the general magnitude of the correlations, rather than the slope in this region. In fact, the slope is another aspect of the correlations we expected from the dependence of neutron and gamma-ray spectra on intrinsic excitation energy. As we pointed out in Section~\ref{sec:the}, inter-fragment correlations give rise to generally positive correlations, since they do not contain direct sources of competition.   

% Qualitatively, the lack of positive correlations in experiment indicates the weakness of inter-fragment correlations. We note also that the positive correlations observed in the calculations are predominantly from continuum to discrete transitions, \textit{i.e.}, the gamma-rays emitted between \textsc{stat} and \textsc{disc}. Due to the difficulty in modeling these transition gamma-rays, we have not attempted to interpret quantitatively the results in this region. 

The correlation enhancement at $E_\gamma = 0.7$ \text{MeV} indicates the existence of positive spin-energy correlations. In fact, the emission of neutrons correlates strongly with the intrinsic excitation energy of the fragment. Enhancements at $E_\gamma \approx 0.7$ MeV indicate that the high-spin, $J \approx 4-8 \  \hbar$ rotational bands are more populated with each neutron emission. This hypothesis is confirmed by comparing \texttt{FIFRELIN} calculations performed with and without spin-energy coupling. In the calculations, both shown in Fig.~\ref{fig:partAn}, we have found that the amplitude of the enhancement is strengthened significantly when the spiin-energy coupling is considered. We have not attempted to quantify the strength of the spin-energy correlations from the experimental data, but this could be pursued in future work. Qualitatively, we can say that the observation of the enhanced correlations indicates the presence of spin-energy correlations in the generation of angular momentum in fission fragments. 

The enhanced correlations at $E_\gamma = 1.2$ MeV reveals less about the spin-energy correlations and more about biasing of the  sample. In fact, the main contributor to this enhancement are low-spin transitions in the spherical fragment region. We note that neutron outgoing energies from near-symmetric fissions are generally higher than for the most probable splits~\cite{Gook2014}. The observation of high-energy neutrons in the lab frame thus biases the fission sample more strongly towards symmetric fission, and is thus positively correlated with high-energy \textsc{disc} gamma-ray emission. 

\subsubsection*{Other Features}

Two other notable features of the energy dependent correlations are their behaviors at the higher and lower end of the gamma-ray spectrum. At high $E_\gamma$, the correlations are most representative of \textsc{stat} emission, since these gamma rays have a harder spectrum than those from \textsc{disc} emission. In the calculations, we find that the intra-fragment contributions to the correlations involving \textsc{stat} gamma rays are predominantly negative. Therefore, we might interpret positive correlations at high $E_n$ and $E_\gamma$ as originating from inter-fragment correlations, and thus conclude from the experiment that inter-fragment correlations are not as strong as intra-fragment correlations. This in turn would lead to the conclusion that the initial states of the fragments, specifically the intrinsic excitation energy, between the two fragments, are not strongly correlated. 

Several problems exist with the above conclusion, which is premature at this stage, and will need to be investigated further. First of all, as mentioned for the enhancement at $1.2$ MeV, gating on higher neutron energies does not only bias towards higher energy fragments, but also towards symmetric fission. Therefore, the correlations shown here are systematically biased by this selection. Second, in the calculations, we find that one of the largest sources of positive correlations in the high $E_\gamma$ region are continuum-discrete transitions. Because these effects cannot be disentangled from the inter-fragment correaltions, we do not attempt to draw conclusions based on this region. 

At low $E_\gamma$, the correlations in the data are more negative than the average. The models predict this feature to originate from inter-fragment correlations, specifically between neutrons from the light fragment and \textsc{disc} gamma rays from the heavy fragment. 

\section{Conclusion}
\label{sec:conc}

We have presented an investigation of the neutron-gamma correlations in \ce{^{252} Cf}(sf). The first result of this study is the development and application of the normalized differential multiplicity covariance, a powerful tool in the analysis of radiation correlations. We have analyzed the correlations between neutrons and gamma rays using the normalized differential multiplicity covariance. This analysis yielded previously unobserved structure in the neutron-gamma correlations. Specifically, we have presented experimental evidence of enhancements in neutron-gamma correlations around the gamma-ray energies of $0.7$ and $1.2$ MeV. 
We have compared the experimental data to model calculations, to determine the physical origins of the observed structure. The appearance of enhancements in the correlations can likely be explained by the correlations between excitation energy and angular momentum. Further theoretical and experimental studies will be needed to quantify the strength of these correlations.

% The comparison of the experiment with the calculations revealed that, contrary to model calculations, the dominant sources of correlations are intra-fragment, \textit{i.e.}, those arising from neutrons and photons emitted from the same fragment. Physically, we have drawn the conclusion that the fragments are not strongly correlated in either energy or angular momentum. This result is in agreement with past experimental and theoretical investigations. Furthermore, the experiment indicates that neutron emission does not change the angular momentum of the fragments by a significant amount, in contrast to the model calculations.  

For nonproliferation applications, the observation of structure in the neutron-gamma correlations represent a further signature of special nuclear material. The presence of positive correlations enhancements at $E_\gamma \approx 0.7$ and $1.2$ MeV, and negative correlations elsewhere, can be exploited by energy-sensitive systems to identify the presence of fissioning material, and distinguish it from other radioactive sources. Preliminary calculations of \ce{^{239} Pu}$(\text{n}_\text{th}, \text{f})$ have shown that the correlation structure is predicted to persists in this reaction as well. 

An experiment involving event-by-event neutron-gamma correlations in coincidence with a fission fragment detector capable of measuring masses and kinetic energies is being planned. Such an experiment would further our understanding of energy and angular momentum correlations in fission and shed light on the decay competition of neutrons and gamma rays near the neutron separation energy. Furthermore, there still remains significant disagreement in the literature about correlations between neutron-gamma competition and fragment properties. The planned experiment will investigate these correlations, perhaps leading to a reconciliation of previous results.

\acknowledgments
S.M. thanks the experimental group at LANSCE-LANL and M.J. Marcath for sharing the experimental data used in this analysis. This work was in part supported by the Office of Defense Nuclear Nonproliferation Research \& Development (DNN R\&D), National Nuclear Security Administration, US Department of Energy. This research was funded in-part by the Consortium for Verification Technology under Department of Energy National Nuclear Security Administration award DE-NA0002534. The work of V.A.P.  was performed under the auspices of UT-Battelle, LLC under Contract No. DE-AC05-00OR22725 with the U.S. Department of Energy. The work of P.T., A.L., and I.S. is carried out under the auspices of the National Nuclear Security Administration of the U.S. Department of Energy at Los Alamos National Laboratory under Contract No. 89233218CNA000001. The work of R.V. was performed under the auspices
of the U.S. Department of Energy by Lawrence Livermore National Laboratory under Contract DE-AC52-
07NA27344. J. R. acknowledges support from the Office of Nuclear Physics in the U.S. Department of Energy under Contract DE-AC02-05CH11231.

\appendix

\bibliography{mybib}

\end{document}